# DEVELOPMENT OF NUCLOTRON MAGNET POWER SYPPLY CONTROL SYSTEM

B. Vasilishin, V. Andreev, V. Gorchenko, A. Kirichenko, A. Kovalenko, I. Kulikov, B. Omelchenko, V. Karpinsky, S. Romanov, B. Sveshnikov, A. Tharenkov, V. Volkov, JINR, Dubna, Russia

## Abstract

The magnetic field control subsystem of the Nuclotron- a superferric heavy ion synchrotron- is described. The bending (BM), focusing (QF) and defocusing (QD) magnets are powered by three supplies. The BMs are driven by the supply of a 6.3 kA nominal current. The QFs and QDs are connected in series and powered by the 6 kA supply. An additional supply of 200 A for the QFs is used to keep the required ratio $I_{QF}/I_{QD}$. The BM magnetic field shape is set by a pulse function generator which produces a reference burst ($B_0$-train) with a 0.1 Gs resolution. A real B-train of the bending magnet field with the same (0.1 Gs) resolution is used for feedback loop and as reference function for the QD and QF supplies. The QD and QF trains are utilized for feedback loop in the QD and QF supplies. The control of slow extraction system supplies is described as well.

## 1 INTRODUCTION

The superconducting synchrotron Nuclotron based on miniature iron-shaped field magnets was put into operation in March 1993 at the Laboratory of High Energies JINR. The accelerator is intended to accelerate nuclei and multi-charged ions including the heaviest ones (uranium) up to an energy of 6 GeV/u for the charge to mass ratio Z/A=1/2. Twenty runs of the accelerator have been performed by the present time. Since December 1999, the slow extraction has operated on the Nuclotron. The Nuclotron Control System (NCS), which is in progress [1], has provided efficient support for a successful operation of the machine during all runs. The most important part of the NCS is the Nuclotron magnet power supply control subsystem (MPSC), which provides control and monitoring of magnetic field in the Nuclotron elements.

There are 96 bending magnets (BM), 32 focusing (QF) and 32 defocusing (QD) quadrupoles, 32 correcting multipoles in the Nuclotron magnetic ring with a circumference of 251.1 m. The maximum value of the magnetic field is about 2 T. The BMs are driven by the supply of a 6.3 kA nominal current. The QFs and QDs are connected in series and excited by the 6.3 kA supply. An additional supply of 200 A for the QFs is used to keep the required ratio $I_{QF}/I_{QD}$ during an accelerator cycle.

The extraction process is realized by excitation of the radial betatron oscillation resonance Qx=20/3. A 20-th harmonic of sextupole nonlinearity is excited by two pairs of sextupoles (ES1…ES4). The lenses of each pair connected in series are supplied from separate sources. It permits one to excite an arbitrary phase of the sextupole disturbance. Four extraction quadrupoles (EQ1…EQ4) connected in series and driven from one source are used to perform the tune shift $\Delta Q_x$ within the resonance band. The beam is extracted by means of an electrostatic septum (ESS) and a Lambertson magnet (LM). The LM is connected in series with the BMs, but there is an additional power supply for the LM used for beam correction in the vertical plane.

## 2 GENERAL MPSC DESCRIPTION

The MPSC performs the following main functionality:

- Downloading and interpreting of information on the required magnetic cycle parameters and calculation of corresponding waveforms.
- Generation of analog signals for control of the Nuclotron magnetic elements in accordance with the parameters specified on the operator console.
- Generation of cycle timing pulses for synchronization of the Nuclotron subsystems with a magnetic cycle.
- Data presentation in graphical and text formats, transmission of the complete set of information on the MPSC status to the database and alarm server.

A schematic diagram of the MPSC is shown in Fig.1. The magnetic cycle is specified at a B(t) level, and the waveforms, which drive the power supplies, are generated by function generators controlled through console software. The BM magnetic field shape is set by a pulse function generator (PFG) which produces 3 reference bursts: $(B_0)^+$-train according to a rising parts of the magnetic field, $(B_0)^-$ -train according to a falling part and $B_0$-train ($B_0=(B_0)^+ + (B_0)^-$). One pulse of the trains corresponds to 0.1 Gs change of the magnetic field. The $(B_0)^+$-train increments and $(B_0)^-$ -train decrements the pattern $B_0$ analog function generator (AFG) based on a 16 bit DAC. Each train consists of a certain number (≤4096) series (vectors) of pulse. The vector is determined by two 24-bit integers: the number of steps (pulses) and step size. Each vector defines part of the magnetic cycle with the constant time derivation, which is determined by the pulse frequency (step size) in this vector. The parabolic form of initial part of the cycle and the edges of flattops is approximated by set of linear parts. The output analog signal ($AB_0$) of the $B_0$ AFG is fed to the magnetic field forming system. The $B_0$-train is passed to the timing pulse selector (TPS) which generates timing pulses for synchronization of Nuclotron subsystems.

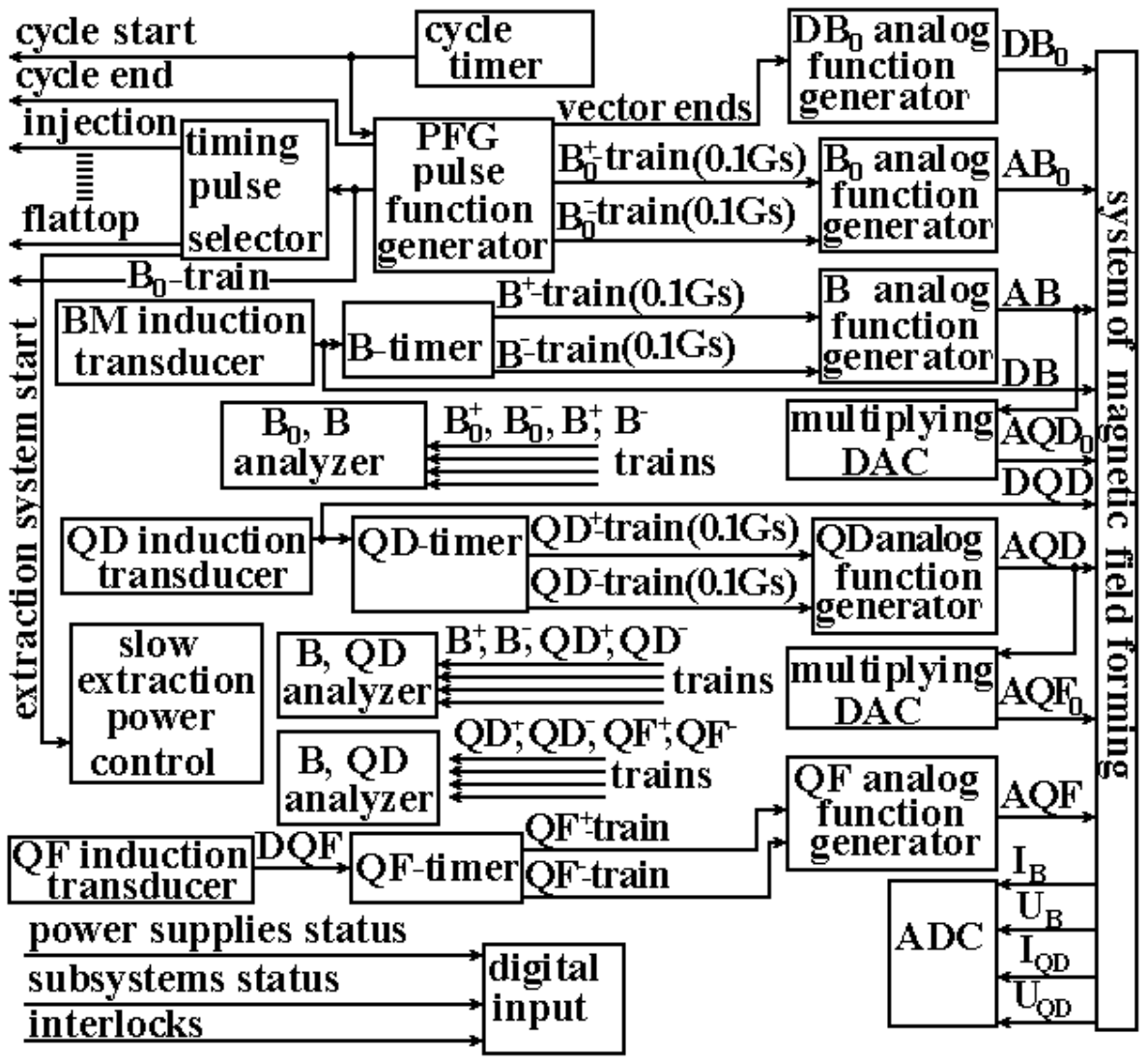

Figure 1: Schematic diagram of the MPSC.

The BM induction transducer produces an analog signal proportional to the derivation (DB) of the real magnetic field in the BMs. This signal enters to the B-timer which generates $B^+$ and $B^-$ trains corresponding to the real magnetic field. These trains are supplied to the B analog function generator, which generates an analog signal AB of the real field in the BMs in the same standard as $AB_0$. The real $B^+$ and $B^-$ trains and the corresponding analog function AB are used for the feedback loop and as a reference for the QD, i.e. the BM power supply is a master and the QD supply is a slave. The QD trains are utilized for control as described above and as a reference for an additional supply of the QFs.

## 3 SLOW EXTRACTION CONTROL

Slow extraction power control starts by the timing pulse from the TPS. A timing diagram of the extraction process is shown in Fig.2. Just after the beginning of a magnetic field cycle, a horizontal orbit bump is created with the help of four correcting dipole magnets (b) in the extraction section to shift the beam to the inner wall of the vacuum chamber. This allows one to avoid beam losses on the ESS and LM during the first turns. After reaching the flattop energy, the horizontal beam tune is shifted to the resonance band edge by decreasing the field gradient in the QDs and QFs (a). At this time, the magnetic fields of the ESs and EQs increase to the set values (e and f). During the extraction, the EQs current is linearly varied (f), and the change rate defines the beam spill duration (h).

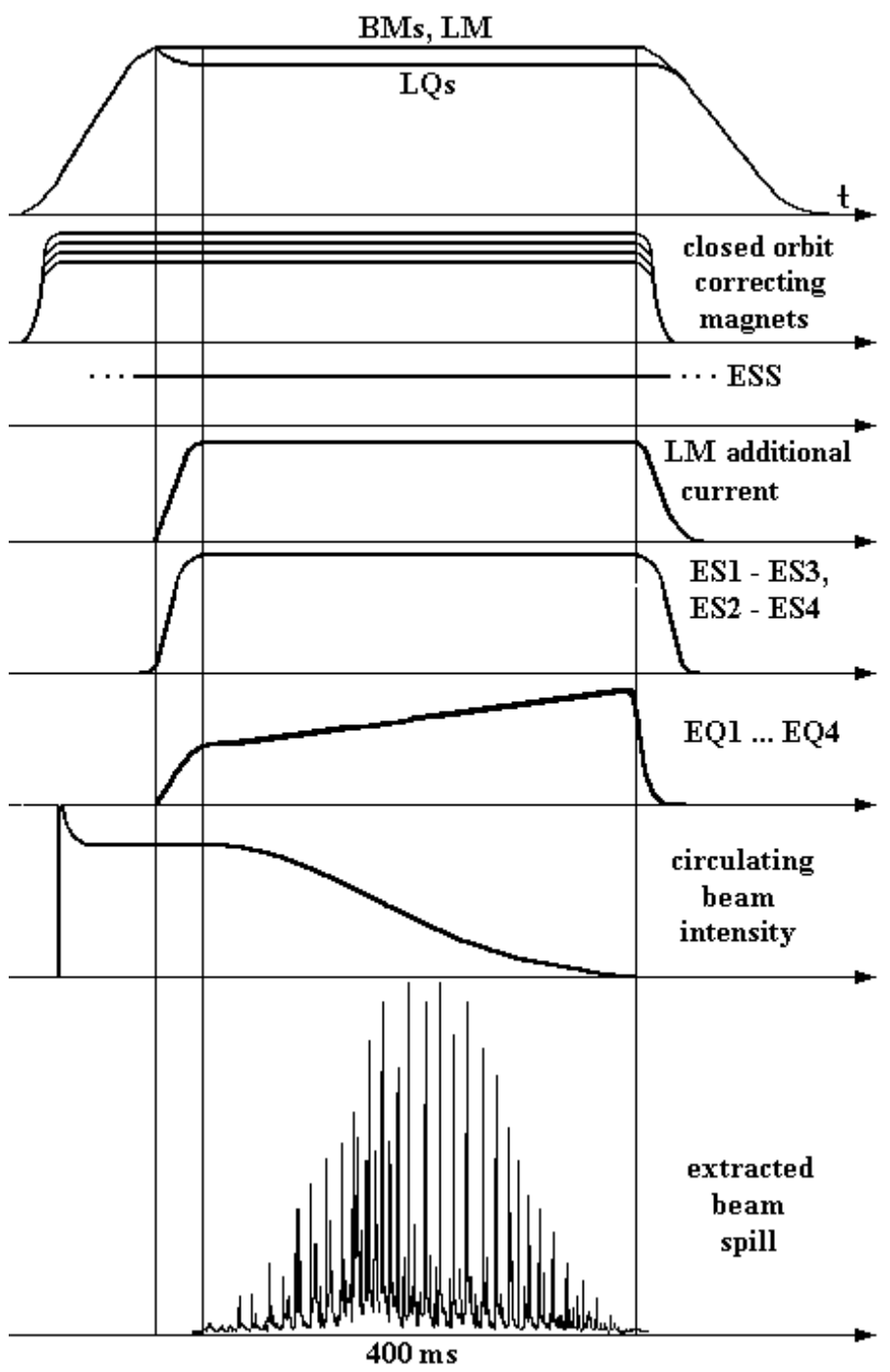

Figure 2: Timing diagram of extraction processes.

The waveforms driving the extraction power supplies are generated by the 16 bit function generators, which are started by a pulse from TPS. All the obtained current waveforms measured with a step of 1 ms are compared with the corresponding references. Difference signals are used for the feedback loop of the corresponding power supplies.

## 4 HARDWARE COMPONENTS

The main part of hardware interface is in the CAMAC standard. All modules were developed and manufactured at LHE, JINR.

A 1 MHz signal generated by the master oscillator of the MPSC is the basic PGF clock. This ensures that the $B_0$-train does not vary by ±0.01Gs.

The PGS has individual outputs for the $B_0$, $(B_0)^+$ and $(B_0)^-$- trains. Pulses from the cycle timer (cycle start) cause the restoration of the initial PGF state and give a start for function generation. The function proceeds until the last defined vector has been produced. The $(B_0)^+$ and $(B_0)^-$-trains, as well as the $B^+$ and $B^-$ -trains, are converted into a 16 bit analog representation ($AB_0$, AB) to control the BM power supply. Pulses of the vector ends are also produced by the PGF. These pulses drive the 16 bit analog function generator of the magnetic field derivation ($DB_0$). In combination with the real field derivations DB and DQD, this function is used to improve dynamic characteristics of the field.

The 2-channel cycle timer generates pulses of the cycle beginning and bursts to drive the measuring devices. The clock period can vary over a range from 1μs to hundreds of seconds. The multi-channel trigger TPS based on a 18 bit counter is used to generate the timing pulse corresponding to a definite value of B.

The digital functions ($B_0$…QF) are monitored using a pulse train analyzer which samples the function at 4096 points during every machine cycle. An additional diagnostics is achieved by a direct measurement of power supply output currents and voltages, with the help of a high performance multi-channel ADC with an onboard memory.

## 5 SOFTWARE

The set of parameters is specified through the console menu. This set includes, for instance, parameters of parabolic segments and linear fractions at various parts of the magnetic cycle, a maximum value of the field, the number of flattops and their duration, flattop locations relative to the magnetic field values, accelerator cycle duration and so on. Software provides the storing and retrieving settings, automatic recording with a time-stamps of all made adjustments, as well as means of stepping back through these changes. The MPSC status is available in the dynamic database. It is updated each accelerator cycle. The archive database keeps a long-term history of the system. The database access (routines) named the Data Viewer allow the users to select data sets for visual observation or printing on the remote workstations (Fig. 3). The alarm server monitors continuously any changes of the MPSC state and detects fault conditions.

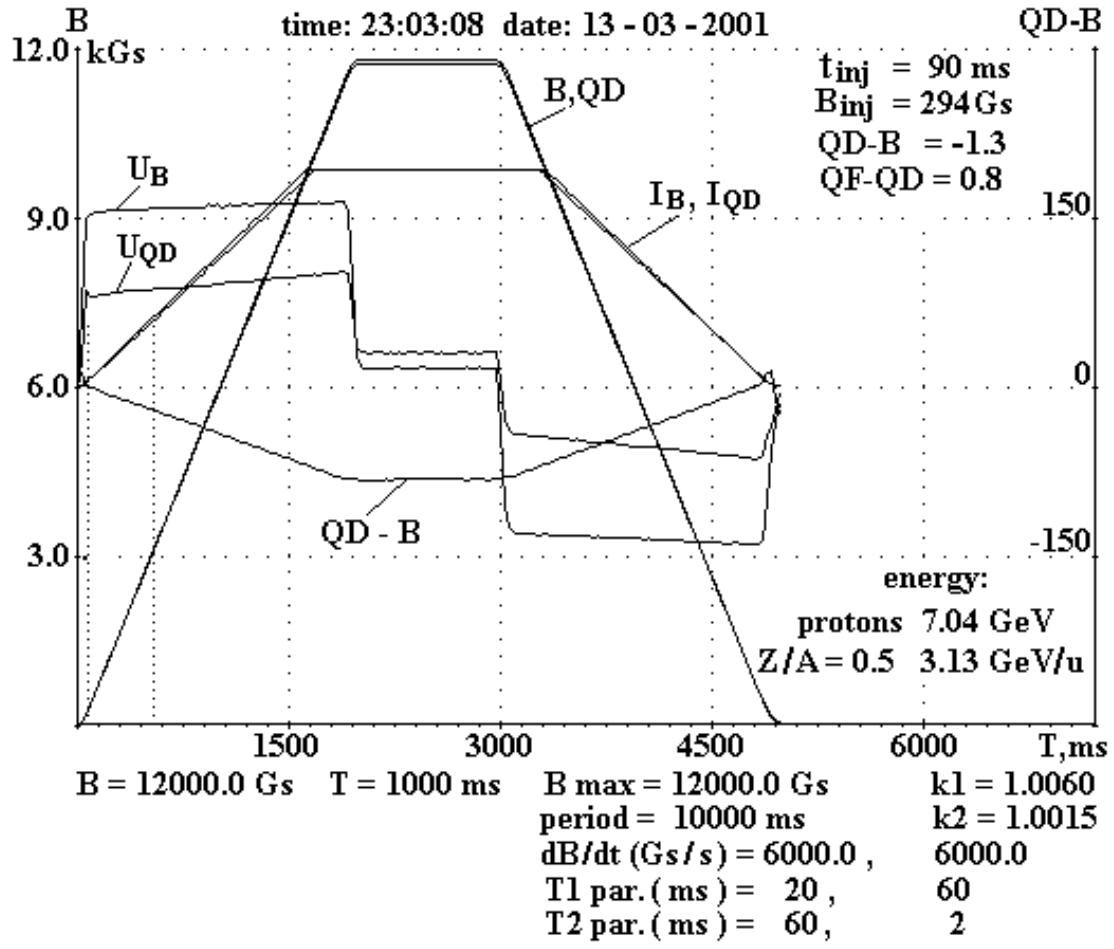

Figure 3: Example of magnetic field cycle parameters.

## 6 CONCLUSION

The MPSC has been successfully used in all Nuclotron runs being capable of generating precision rumps ranging from a fraction of kGs/s up to 10kGs/s. Modifications will be undertaken to improve the stability of slow extraction supplies and to reduce the ripples in the extracted beam. The 16 bit DACs will be replaced with 18 bit devices. These improvements will permit one to realize the feedback loop “extracted beam current→power supplies”.

The authors are grateful to L.Sveshnikova for her help in preparing this paper.